\documentclass[useAMS,usenatbib]{mn2e}

\title[TeV neutrinos and gamma rays from pulsars]{TeV neutrinos and gamma rays from pulsars}
\author[A. Bhadra and R. K. Dey]{A. Bhadra$^{1}$\thanks{E-mail:
aru\_bhadra@yahoo.com} and R. K. Dey
$^{2}$\thanks{rkdey2007@rediffmail.com}\\
$^{1}$High Energy \& Cosmic Ray Research Centre, University of North Bengal, Siliguri, WB 734013 INDIA\\
$^{2}$Department of Physics, University of North Bengal, Siliguri, WB 734013 
INDIA}

\begin{document}

\date{Accepted . Received ; in original form }

\pagerange{\pageref{firstpage}--\pageref{lastpage}} \pubyear{}

\maketitle

\label{firstpage}

\begin{abstract}
Recent studies suggest that pulsars could be strong sources of TeV muon neutrinos provided positive ions are accelerated by pulsar polar caps to PeV energies. In such a situation muon neutrinos are produced through the delta resonance in interactions of pulsar accelerated ions with its thermal radiation field. High energy gamma rays also should be produced simultaneously in pulsar environment as both charged and neutral pions are generated in the interactions of energetic hadrons with the ambient photon fields. Here we estimate TeV gamma ray flux at Earth from few nearby young pulsars. When compared with the observations we find that proper consideration of the effect of polar cap geometry in flux calculation is important. Incorporating such an effect we obtain the (revised) event rates at Earth due to few potential nearby pulsars. The results suggest that pulsars are unlikely to be detected by the upcoming neutrino telescopes. We also estimate TeV gamma ray and neutrino fluxes from pulsar nebulae for the adopted model of particle acceleration.
\end{abstract}

\begin{keywords}
neutrinos, gamma rays:theory, pulsars:general.
\end{keywords}

\section{Introduction}

Probable candidates of high energy neutrino radiation include Gamma ray bursts, Active galactic nuclei etc (Halzen \& Hooper 2002). Recently Link and Burgio (in this article we hereafter refer them as LB) (Link \& Burgio 2005, 2006) have shown that pulsars also could be strong source of high energy neutrinos. As per their results pulsars with a magnetic moment component antiparallel to the spin axis, which is expected in half of total neutron stars, could emit TeV muon neutrinos with fluxes observable by the operating or planed large area neutrino observatories. As a conjecture to be verified by observations LB considered that protons or heavier ions are accelerated near the surface of the pulsar by the polar caps to PeV energies. When these accelerated ions interact with the thermal radiation field of pulsar the $\Delta$ resonance state may occur provided their energies exceed the threshold energy for the process. Though radiation losses limit the maximum energy that can be attained by a nuclei in the acceleration process but such an energy condition is expected to satisfy for several pulsars. Muon neutrinos are subsequently produced from the decay of $\Delta$ particles. 

The presence of a hadronic component in the flux of pulsar accelerated particles should result in emission of high energy neutrinos and gamma rays simultaneously as both charged and neutral pions are produced in the interactions of energetic hadrons with the ambient photon fields surroundings the acceleration region. Here we would show that for the model adopted by LB, the estimated TeV gamma ray fluxes from several nearby pulsars are higher than the observed upper limits of fluxes. In the quest for reasons of such inconsistency we note that an implicit assumption in the LB estimation of neutrino flux is that the polar cap area is equal to the neutron star surface area. Such an assumption seems unreasonable in view of the polar cap geometry (Beskin, Gurevich \& Istomin 1993). When this fact is taken into consideration, the stated inconsistency between the estimated and the observed gamma ray fluxes is found to disappear. In view of this observation a revised estimates of the neutrino fluxes from few known gamma ray pulsars are obtained incorporating polar cap geometry. 

A young neutron star is generally encircled by pulsar wind nebula. Positive ions, after gaining energy from polar gaps will move away from the pulsar practically along the open field lines and will finally inject into the nebula. It is very likely that these energetic ions would be trapped by the magnetic field of the nebula for a long period and consequently they should produce high energy gamma rays and neutrinos by interacting with the matter of the nebula. We, therefore, estimate the expected flux of TeV gamma rays from pulsar nebulae and by comparing with the observation for a couple of well known nebulae we check the consistency of the model. 
We also calculate the flux of TeV neutrinos from couple of pulsar nebulae.    

The organization of the article is as follows. In the next section after describing generation of the TeV gamma rays and neutrinos in a pulsar environment, the expected flux of gamma rays from a pulsar is estimated for the polar cap model of acceleration as used by LB. By comparing the model predicted $\gamma$ ray fluxes from some potential young pulsars with the  observations, importance of inclusion of polar cap area in the calculation has been stressed in section 3. Incorporating such a feature (revised) event rates in a neutrino telescope at Earth due to few potential pulsars are obtained in section 4. In section 5, fluxes of TeV gamma rays and neutrinos from pulsar nebulae are obtained for the adopted model and finally the results are concluded in section 6.  

\section{TeV gamma rays and neutrinos from pulsars}
Several detailed mechanisms  have so far been suggested for acceleration of particles by pulsars those include the popular polar gap (Ruderman \$ Sutherland 1975; Arons \& Scharlemann 1979;  Daugherty \& Harding 1996; Harding \& Muslimov 1998) and the outer gap models (Cheng, Ho \& Ruderman, 1986). In the former model, acceleration of particles takes place in the open field line region above the magnetic pole of the neutron star whereas in the case of outer gap model it occurs in the vacuum gaps between the neutral line and the last open line in the magnetosphere. Thus, the region of acceleration in the polar gap model is close to the pulsar surface, while the same in the outer gap model is close to the light cylinder. 

In the polar cap acceleration model, particles are extracted from the polar cap and accelerated by large rotation-induced electric fields, forming the primary beam. The maximum potential drop that may be induced across the magnetic field lines between the magnetic pole and the last field lines that opens to infinity is $\Delta \phi = B_{S}R^{3}\Omega^{2}/2c^{2}$ (Goldreich \& Julian, 1969) where $B_{S}$ is the strength of magnetic field at neutron star surface, $R$ is the radius of the neutron star, $\Omega$ is the angular velocity and $c$ is the speed of light. Accordingly, for young millisecond pulsar with high magnetic fields ($B_{S} \sim 10^{12}$ Gauss), the magnitude of the potential drop could be as large as $7 \times 10^{18} B_{12} P_{ms}^{-2}$ volts ($B_{S} \equiv B_{12} \times 10^{12}$ Gauss). But in young pulsars, the electric field along magnetic field lines is likely to be screened well below the vacuum potentials by the onset of electron-positron pair cascades in the strong magnetic field (Cheng \& Ruderman 1977) and this would limit the potential to about $10^{12}$ eV. Though the flux of synchrotron radiation observed from the Crab and other pulsar wind nebula (PWN) indiactes such a possibility but questions like where in the magnetospher are pairs created and how many are created, are still not settled and one cannot rule out the possibility that ions can reach energies equal to a significant part of the total potential drop through polar cap acceleration, particularly in view of the observation of gamma radiation of energies above tens and even hundreds of TeV from Crab and other PWN. 

LB (Link \& Burgio 2005, 2006) conjectured that protons or heavier ions are accelerated near the surface of a pulsar by the polar caps to PeV energies (corresponds to no/little screening) when $\bf{\mu . \Omega} <0$ (such a condition is expected to hold for half of the total pulsars). When pulsar accelerated ions interact with the thermal radiation field of pulsar the $\Delta$ resonance state may occur provided their energies exceed the threshold energy for the process. The threshold condition for production of $\Delta$ resonance state in $p\gamma$ interaction is given by

\begin{equation}
\epsilon_{p}\epsilon_{\gamma} (1-cos\theta_{p\gamma}) \ge 0.3 \; GeV^{2} 
\end{equation} 

where $\epsilon_{p}$ and $\epsilon_{\gamma}$ are the proton and photon energies respectively and $\theta_{p\gamma}$ is the incident angle between the proton and photon in the laboratory frame. In a young neutron star with surface temperature $T_{\infty}$ the energy of a thermal photon near the surface of the neutron star is $2.8kT_{\infty}(1+z_{g})$, $z_{g}$ ($\sim 0.4$) being the gravitational redshift. This implies that in a young pulsar atmosphere the condition for production of the $\Delta$ resonance is $B_{12} P_{ms}^{-2}T_{0.1 \; keV} \ge 3 \times 10^{-4}$ (Link \& Burgio 2005, 2006) where $T_{0.1 \; keV} \equiv (kT_{\infty}/0.1 keV)$, $T_{\infty} \sim 0.1 keV$ being the typical surface temperature of young pulsars. Such a condition holds for many young pulsars and thus $\Delta$ resonance should be reached in pulsar atmosphere. Gamma rays and neutrinos are subsequently produced through the following channels:

\[p + \gamma \rightarrow \Delta^{+} \rightarrow \left\{\begin{array}{ll}
p + \pi^{o} \rightarrow p + 2\gamma \\
n\pi^{+} \rightarrow n + e^{+} + \nu_{e} + \nu_{\mu} + \bar{\nu_{\mu}}
\end{array}
\right. \] 
 
Since the charge-changing reaction takes place just one-third of the time, on the average  four high-energy gamma-rays are produced for every three high-energy neutrinos (or for every two high energy muon neutrino and antineutrino) when a large number of such reactions occur.

The flux of gamma rays and muon neutrinos from pulsars can be estimated as follows: The charge density of ions near the pulsar surface is $\rho_{q} \simeq eZn_{o}$, where $n_{o}(r)\equiv B_{s}R^{3} \Omega /(4 \pi Ze c r^{3})$ is the Goldreich-Julian density (Goldreich \& Julian, 1969) of ions at radial distance $r$. For acceleration to take place there must be a charge depleted gap (here polar gap) and the densiy in the gap may be written as $f_{d}(1-f_{d}) n_{o}$, where $f_{d}$ ($ < 1$) is the depletion factor which is a model dependent quantity ($f_{d}=0$ corresponds to zero depletion). The flux of protons accelerated by a polar cap is therefore 

\begin{equation}
I_{pc}=c f_{d}(1-f_{d}) n_{o} A_{pc}, 
\end{equation}

where $A_{pc}=\eta 4 \pi R^{2}$ is the area of the polar cap, $\eta$ is tha ratio of polar cap area to the neutron star surface area, which is taken as unity by LB in their work (i.e. in original calculation by LB $A_{pc}$ was taken as equal to the surface area of the whole neutron star). The canonical polar cap radius is given by $r_{pc}= R (\Omega R/c)^{1/2}$ (Beskin et~al.\ 1993) and thus 

\begin{equation}
\eta = \Omega R/(4c).
\end{equation}

The protons accelerated by polar cap will interact with the thermal radiation field of the neutron star. The temperature of polar caps is expectedly higher than the surface temperature of neutron star but the contribution of polar caps on the thermal radiation field of a neutron star should be negligible because of their small surface area in comparison with the surface area of the whole neutron star. For a young neutron star with surface temperature $T_{\infty}$ the photon density close to the neutron star surface is $n_{\gamma}(R)=(a/2.8k)\left[(1+z_{g})T_{\infty}\right]^{3} $, $a$ being the Stefan-Boltzmann constant. Numerically $n_{\gamma}(R)\simeq 9 \times 10^{19} T^{3}_{0.1 \; keV}$. At radial distance $r$, the photon density will be $n_{\gamma}(r)=n_{\gamma}(R)(R/r)^{2}$. The probability that a PeV energy proton starting from the pulsar surface will produce $\Delta$ particle by interacting with thermal field is given by (Link \& Burgio 2005) $P_{c}=1-\int_{R}^{r} P(r)$ where $dP/P =-n_{\gamma}(r) \sigma_{p\gamma} dr$ . The threshold energy for production of $\Delta$ resonance state in $p\gamma$ interaction as given by Eq.(1) increases rapidly with distance from the surface of neutron star because of the $(1-cos\theta_{p\gamma})^{-1}$ factor (Link \& Burgio 2005). Requiring conversion to take take place in the range $R \le r \le 1.2R$ (at $r=1.2R$ the value of $(1-cos\theta_{p\gamma})^{-1}$ averaged over surface becomes double), $ P_{c}$ has been found as $\simeq 0.02 T^{3}_{0.1 \; kEV}$ (Link \& Burgio 2005). From a follow up calculation by Link \& Burgio (2006) it is found that conversion probability is slightly lower than that mentioned above and can be parameterized as $P_{c} \simeq  T^{3}_{0.1 \; kEV}$. Thus the total flux of neutrino/gamma ray generated in pulsar from the decay of $\Delta^{+}$ resonance is

\begin{equation}  
I =  2 c \xi A_{pc} f_{d} (1-f_{d}) n_{o} P_{c}
\end{equation}

where $\xi$ is $4/3$ and $2/3$ for gamma rays and muon neutrinos respectively. Denoting the duty cycle of the gamma ray/neutrino beam as $f_{b}$ (typically $f_{b} \sim 0.1 -0.3$), the phase averaged gamma ray/neutrino flux at Earth from a pulsar of distance d is given by 

\begin{equation}
\phi \simeq  2 c \xi \zeta \eta f_{b} f_{d} (1-f_{d}) n_{o}\left(\frac{R}{d}\right)^{2}P_{c}
\end{equation} 

where $\zeta$ represents the effect due to neutrino oscillation (the decays of pions and their muon daughters results in initial flavor ratios $\phi_{\nu_{e}}:\phi_{\nu_{\mu}}:\phi_{\nu_{\tau}}$ of nearly $1:2:0$ but at large distance from the source the flavor ratios is expected to become $1:1:1$ due to maximal mixing of $\nu_{\mu}$ and $\nu_{\tau}$). $\zeta =1$ and $1/2$ for gamma rays and muon neutrinos respectively.  

Average energy of the produced muon neutrinos would be $\epsilon_{\nu_{\mu}} \sim 50 T^{-1}_{0.1 \; keV}$ TeV (Link \& Burgio 2005, 2006) whereas that for gamma rays is expected to be $E_{\gamma} \sim 100 T^{-1}_{0.1 \; keV}$ TeV. 

\section{TeV gamma rays from few potential pulsars: Comparison with observations}

Though there are about $1800$ pulsars known through radio detections, only few have been detected in the gamma-rays. From observations made with gamma ray telescopes on satellites so far only seven high confidence gamma-ray pulsars are known in the energy range up to few GeV (Thompson 2003). Besides, three other pulsars (PSR B1046-58, B0656+14, J0218+4232) are considered to be gamma-ray emitters with a lower confidence level (Thompson 2003) in the same energy region. Some basic properties of the high confidence {\it young} gamma-ray pulsars, namely distance (D), spin-down age, period (P), the calculated magnetic field strength at the neutron star surface, surface temperature and pulsar duty cycle ($f_{b}$) are listed in the Table 1. 

\begin{table*}
\begin{center}
\begin{tabular}
{|l|l|l|l|l|l|r|} \hline
     Source         & {\em d}   & {\em age}      &{\em p}      & {\em $B_{12}$}    & {\em $T_{0.1 keV}$}   & $f_{b}$   \\
                    &    kpc    &    year        &    ms       &                   &                      &         \\
\\ \hline

      Crab    &   2             & $10^{3}$       &   33       &      3.8           & $\sim 1.7$             &   0.14      \\ 
&&& \\ \hline

     Vela     &  0.29           &   $10^{4.2}$    &   89       &      3.4          &   0.6               &   0.04      \\ 
&&& \\ \hline

$B 1706-44$   &  1.8       &  $10^{4.3}$     & 102       &       3.1               & 1  ?                 &   0.13       \\ 
&&& \\ \hline 

$B 1509-58$   &  4.4       &   $10^{3.2}$    & 151       &      0.26               & 1    ?               &   0.26  \\ 
&&& \\ \hline 

$J 0205+64$   &  3.2       &   $10^{2.9}$    & 65         &    3.8                 &   0.04                & 0.9      \\ 
&&& \\ \hline

\end{tabular}
\caption {The characteristics of high confidence young low energy gamma ray pulsars} 
\end{center}
\end{table*}

None of the listed pulsars are, however, detected at TeV energies despite recent great improvement in the knowledge of Galactic gamma-ray sky above 100 GeV mostly by means of ground-based Imaging Atmospheric Cherenkov telescope systems such as the CANGAROO, HEGRA , High Energy Stereoscopic System (HESS) or the Major Atmospheric Gamma-ray Imaging Cherenkov Observatory (MAGIC).  Up to now no evidence for pulsed emission from any other pulsar has been found from the observations (Chadwick et~al.\ 2000; de Naurois et~al.\ 2002; Lessard et~al.\ 2000; Aharonian et~al.\ 2004; Aharonian et~al.\ 2007), and only upper limits on the pulsed VHE gamma-ray flux are derived under various assumptions on the characteristics of the pulsed emission. For the pulsars listed in the table 1 the observed upper limits of integral fluxes  are given in the table 2. The upper limits are obtained from the differential spectra assuming a power law differential spectrum with index $-2.4$. The upper limits of integral fluxes of gamma rays from the pulsars at around hundred TeV as obtained by the extensive air shower experiments such as the Tibet (Amenomori et~al.\ 2008; Wang et~al.\ 2008) are found less restrictive.  

Assuming pulsar accelerated ions are protons, numerical values of the integral TeV gamma ray fluxes are obtained for the pulsars listed in Table 2 from Eq.(5) and are also shown in the same Table (for the numerical estimation of flux we take $Z=1$, and $f_{d}=1/2$ throughout the present work).   

\begin{table*}
\begin{center}
\begin{tabular}
{|l|l|l|r|} \hline
Source & \multicolumn{2}{c|} {\em Expected integral flux }  & {\em Observed upper limit of integral flux}  \\ \\ \cline{2-3}  

       &  $\eta= 1$     &  $\eta$ as given by Eq.(3)         &         \\ \\

       & $ 10^{-15} cm^{-2}s^{-1}$ &  $10^{-15} cm^{-2}s^{-1}$ &  $10^{-15} cm^{-2}s^{-1}$ \\  \\ \hline

Crab   &  1012 & 1.65 &  8 (56)    \\ 
&&& \\ \hline

Vela  & 208 & 0.123 &   10 (20)  \\ 
&&& \\ \hline

$B 1509-58$  & 67 & .0034 & 10 (20)  \\ 
&&& \\ \hline

$B 1706-44$  & 71 & .0025 & 10 (20) \\ 
&&& \\ \hline 

\end{tabular}
\caption {Comparison of predicted versus observed integral TeV gamma ray fluxes from some nearby young gamma ray pulsars. The numbers in parentheses are the energy thresholds in TeV for which upper limits are determined. The observed upper limits for Crab pulsar is due to \citet{ab4} whereas for the rest of the pulsars the observed upper limits are due to \citet{ab6}}. 
\end{center}
\end{table*}

It is clear from the Table 2 that the model with $\eta =1$ is not consistent with the observed upper limits; the observed limits are substantially lower than the predicted fluxes. The observations made so far, however, do not rule out the model with $\eta$ given by Eq.(3). 

\section{TeV neutrinos from pulsars}
When $\eta$ is given by Eq.(3), neutrino fluxes from nearby young gamma ray pulsars would be much lower than estimated by LB (Link \& Burgio 2006). A high energy muon neutrino is usually detected indirectly, through observation of the secondary high energy muon produced by the muon neutrino on interaction with the ice or rock in the vicinity of a neutrino telescope via charged current interactions. The track of the produced muon is usually reconstructed by detecting the Cerenkov light emitted by the muon as it propagates through the telescope. The probability of detection of muon neutrinos is the product of the interaction probability of neutrinos and the range of the muon and is $
p_{\nu \rightarrow \mu} \simeq 1.3 \times 10^{-6} \left( \epsilon_{\nu_{\mu}}/ 1 \; TeV \right) 
$ (Gaisser et~al.\ 1995). The expected event rates in a neutrino telescope due to potential young pulsars are given in Table 3.
 
\begin{table*}
\begin{center}
\begin{tabular}
{|l|r|} \hline
Source &  {\em Expected event rates } \\ \\

       & $ km^{-2}yr^{-1}$  \\  \\ \hline

Crab   &  0.008  \\ 
& \\ \hline

Vela  & .0006  \\ 
& \\ \hline

$B 1509-58$  & .00016  \\ 
& \\ \hline

$B 1706-44$  & .00012 \\ 
& \\ \hline 

\end{tabular}
\caption {Expected event rates in a neutrino telescope due to some nearby young gamma ray pulsars. } 
\end{center}
\end{table*}

The event rates are clearly very low and thus possibility of observing pulsars by a kilometer scale neutrino detector does not look bright. 

\section{Gamma rays and neutrinos from pulsar nebula}

A young neutron star is generally encircled by pulsar wind nebula. Positive ions, after gaining energy from polar gaps will move away from the pulsar practically along the open field lines and will finally inject into the nebula. It is very likely that these energetic ions would be trapped by the magnetic field of the nebula for a long period and consequently they should produce an appreciable of high energy gamma rays/neutrinos by interacting with the matter of the nebula.

\subsection{Magnetic trapping of pulsar accelerated PeV ions in nebulae} 
Conservation of magnetic flux across the light cylinder entails that outside the light cylinder $B \sim r^{-1}$ whereas far from the light cylinder radial component of magnetic field varies as $B_{r} \sim r^{-2}$. Thus (far) outside the light cylinder the azimuthal component of the magnetic field dominates over the radial field. Therefore, accelerated protons while moving away from the pulsar have to cross the field lines (for instance magnetic field lines at wind shock). The Larmor radius of particles (even for proton) of energy about $1$ PeV is expected to be smaller than the radius of nebula during most of the time of the evolution of nebula (Bednarek  \& Protheroe 2002; Bhadra 2006). Thus it is very likely that energetic particles of PeV energies would be trapped by the magnetic field of the nebula. The energetic particles propagate diffusively in the envelope and they escape from nebula when the mean radial distance traveled by the particles becomes comparable with the radius of nebula at the time of escaping. This time is somewhat uncertain due to uncertainty of the value of diffusion coefficient but is estimated as at least few thousand years (Bednarek \& Protheroe 2002; Bhadra 2006).

\subsection{Gamma rays and neutrinos from nebulae of young pulsars}
As pointed out in the preceding section the pulsar injected ions of PeV energies should be trapped by the magnetic field of the nebula for a long period and consequently there would be an accumulation of energetic ions in nebula. These energetic ions will interact with the matter of the nebula. The rate of interactions $(\xi)$ would be $ n c \sigma_{pA}$, where $n $ is the number density of protons in nebula and $\sigma_{pA}$ is interaction cross-section. In each such interaction charged and neutral pions will be produced copiously. Subsequently the decays of neutral pions will produce gamma rays whereas charged pions and their muon daughters will give rise to neutrinos.   

If $m$ is the mean multiplicity of charged particles in proton-ion interaction, then the flux of gamma rays at a distance $d$ from the source roughly would be

\begin{equation} 
\phi_{\gamma} \approx 2 c \beta \eta f_{d} (1-f_{d}) n_{o}\left(\frac{R}{d}\right)^{2}\xi m t
\end{equation}

where $\beta$ represents the fraction of pulsar accelerated protons trapped in the nebula and $t$ is the age of the pulsar. Note that there should not be any reduction of flux due to pulsar duty cycle in the case of emission to nebula. Though $n_{o}$ is taken as constant but actually at the early stages of pulsar $n_{o}$ should be larger owing to the smaller pulsar period. So the above expression gives only a lower limit of flux. Typical energy of these resultant gamma rays would be $\sim 10^{3}/(6*m)$ TeV where for (laboratory) collision energy of $1$ Pev m is about $32$ (Alner et~al.\ 1987).  

Numerical values of the integral TeV gamma ray fluxes from two nearby nebulae, Crab nebula and Vela nebula have been estimated for perfect trapping of pulsar accelerated protons in nebulae and are shown in Table 4. The observed integral gamma ray fluxes above 1 TeV from Crab nebula (Aharonian et~al.\ 2006a) and Vela nebual (Aharonian et~al.\ 2006b) are also given there for comparison. 

\begin{table*}
\begin{center}
\begin{tabular}
{|l|l|l|r|} \hline
PWNe & {\em n} & {\em Expected flux} & {\em Observed flux} \\
& $cm^{-3}$ &    $\times 10^{-12} cm^{-2}s^{-1}$   & $10^{-12} cm^{-2}s^{-1}$ \\
\\ \hline
Crab nebula  & 150  & 0.6  & 22.6   \\ 
&&& \\ \hline
Vela nebula & 1 & 0.4 & 12.8   \\ 
 &&& \\ \hline
\end{tabular}
\caption {Comparison of predicted versus observed integral gamma ray fluxes from two nearby pulsar wind nebulae} 
\end{center}
\end{table*}

The neutrino fluxes from the nebulae would be of nearly the same to those of gamma rays. Incorporating the neutrino oscillation effect the expected event rates in a neutrino telescope due to TeV muon neutrinos from nebulae of Crab and Vela are $0.2 \;km^{-2}yr^{-1}$ and $0.1 \;km^{-2}yr^{-1}$ respectively. 
Note that the event rates obtained here are rough numerical values. The flux will be higher if the accelerated ion is heavier than proton.      

\section{Conclusion}

To summarize, the present work suggests that pulsars are unlikely to be strong sources of TeV neutrinos. The non detection of any statistically significant excess from the direction of any pulsar by the AMANDA-II telescope (Ackermann et~al.\ 2008; Ackermann et~al.\ 2005; Ahrens et~al.\ 2004) is thus as per expectations. 

If protons are accelerated to PeV energies by the pulsar, then pulsar nebulae are more probable sites of energetic neutrinos provided energetic particles of PeV energies are efficiently trapped by the magnetic field of the nebulae. But even for pulsar nebulae the expected event rates are small and the detection probability of pulsar nebulae by the upcoming neutrino telescopes, such as the IceCube (Halzen 2006), is very low.       

\section*{Acknowledgments}

The authors would like to thank an anonymous referee for very useful comments and suggestions.


\begin{thebibliography}{99}

\bibitem[\protect\citeauthoryear{Ackermann et al.}{2005}]{ab1} Ackermann M. et al., 2005, Phys. Rev. D 71, 077102

\bibitem[\protect\citeauthoryear{Ackermann et al.}{2008}]{ab2} Ackermann M. et al., 2008, APJ, 675, 1014

\bibitem[\protect\citeauthoryear{Aharonian et al.}{2004}] {ab3} Aharonian F. et al. (HEGRA collaboration), 2004, ApJ, 614, 897 

\bibitem[\protect\citeauthoryear{Aharonian et al.}{2006a}] {ab4} Aharonian F. et al., (HESS Collaboration) 2006a, Astron. Astrophys. 457, 899 

\bibitem[\protect\citeauthoryear{Aharonian et al.}{2006b}] {ab5} Aharonian F. et al., (HESS Collaboration) 2006b, Astron. Astrophys. 448, 206 

\bibitem[\protect\citeauthoryear{Aharonian et al.}{2007}] {ab6} Aharonian F. et al., (HESS collaboration) 2007, Astron. Astrophys. 466, 543

\bibitem[\protect\citeauthoryear{Ahrens et al.}{2004}] {ab7} Ahrens J. et al., 2004, Phys. Rev. Lett. 92, 071102  

\bibitem[\protect\citeauthoryear{Alner et al.}{1987}] {ab8} Alner, G. J. et al., 1987, Phys. Rep. 154, 247 

\bibitem[\protect\citeauthoryear{Amenomori et al.}{2008}] {ab9} Amenomori M. et al. (Tibet ASgamma Collaboration), 2008, Nucl. Phys. Proc. Suppl. 175-176: 431

\bibitem[\protect\citeauthoryear{Arons \& Scharlemann }{1979}] {ab10} Arons J. and  Scharlemann E.~T., 1979, ApJ, 231, 854

\bibitem[\protect\citeauthoryear{Bednarek \& Protheroe}{2002}] {ab11} Bednarek W. and Protheroe R. ~J., 2002, 16, 397

\bibitem[\protect\citeauthoryear{Beskin et al.}{1993}] {ab12} Beskin V. S., Gurevich A. V. and Istomin N. Ya., 1993, Physics of the Pulsar Magnetosphere (Cambridge University Press) 117

\bibitem[\protect\citeauthoryear{Bhadra}{2006}] {ab13} Bhadra A., 2006, Astropart. Phys. 25, 226 

\bibitem[\protect\citeauthoryear{Cheng \& Rederman}{1977}] {ab15} Cheng K.~S., and Ruderman M., 1977 ApJ, 214, 598 

\bibitem[\protect\citeauthoryear{Chadwick et al.}{2000}] {ab14} Chadwick P. M. et al., 2000, ApJ, 537, 414

\bibitem[\protect\citeauthoryear{Cheng et al.}{1986}] {ab15} Cheng K.~S., Ho C. and Ruderman M., 1986 ApJ, 300, 500 

\bibitem[\protect\citeauthoryear{Daugherty \& Harding}{1996}] {ab16} Daugherty J. K. and Harding A. K. , 1996, ApJ, 458, 278

\bibitem[\protect\citeauthoryear{de Naurois et al.}{2002}] {ab17} de Naurois M. et al., 2002, ApJ,566, 343

\bibitem[\protect\citeauthoryear{gaisser et al.}{1995}] {ab18} Gaisser T. K., Halzen F., and Stanev T., 1995, Phys. Rept., 258, 173

\bibitem[\protect\citeauthoryear{Goldreich \& Julian}{1969}] {ab19} Goldreich P. and Julian W.~H., 1969, ApJ, 157, 869 

\bibitem[\protect\citeauthoryear{(Halzen \& Hooper)} {2002}] {ab20} Halzen F. and Hooper D.,  2002, Prog. Theor. Phys. 65, 1025 

\bibitem[\protect\citeauthoryear{Halzen}{2006}] {ab21} Halzen F., 2006, Eur. Phys. J. C 46, 669

\bibitem[\protect\citeauthoryear{Harding \& Muslimov}{1998}] {ab22} Harding A.~K. and Muslimov A.~G., 1998, ApJ, 508, 328

\bibitem[\protect\citeauthoryear{Lessard et al.}{2000}] {ab23} Lessard R. W. et al., 2000, ApJ, 531, 942 

\bibitem[\protect\citeauthoryear{Link \& Burgio}{2006}] {ab24} Link B. and  Burgio F., 2006 MNRAS 371, 375 

\bibitem[\protect\citeauthoryear{Link \& Burgio}{2005}]{ab25} Link B. and Burgio F., 2005, Phys. Rev. Letts., 94, 181101 

\bibitem[\protect\citeauthoryear{Ruderman \& Sutherland}{1975}] {ab26} Ruderman M.~A. and Sutherland P.~G., 1975 ApJ, 196, 51

\bibitem[\protect\citeauthoryear{Thompson}{2003}] {ab27} Thompson D. J., 2003, astro-ph/0312272 

\bibitem[\protect\citeauthoryear{Wang et al.}{2008}] {ab28} Wang Y. et al., 2008, astro-ph/ 08041862  

\end{thebibliography}
\end{document}